\documentclass[final,5p,times,twocolumn]{elsarticle}
\usepackage{graphicx}
\usepackage{amsmath}
\usepackage{amssymb}
\usepackage{bbm,array}

\journal{Physica E}

\begin{document}

\newcommand{\be}{\begin{equation}}
\newcommand{\beq}{\begin{equation}}
\newcommand{\ee}{\end{equation}}
\newcommand{\bea}{\begin{eqnarray}}
\newcommand{\eea}{\end{eqnarray}}
\newcommand{\ba}{\begin{array}}
\newcommand{\ea}{\end{array}}
\newcommand{\Id}[1] {\int \! \! {\rm d}^3 #1}
\newcommand{\ID}[1] {\int \! \! \! \frac{{\rm d}^3 #1}{(2 \pi)^3}}
\newcommand{\citenew}[1] {\refnote{\cite{#1}}}
\newcommand{\w}{\omega}
\newcommand{\g}{\gamma}
\newcommand{\G}{\Gamma}
\newcommand{\vr} {{\bf r}}
\newcommand{\vj} {{\bf j}}
\newcommand{\vp} {{\bf p}}
\newcommand{\vs} {{\bf s}}
\newcommand{\nup}{n_{\uparrow}}
\newcommand{\ndown}{n_{\downarrow}}
\def\v#1{\mbox{\boldmath $#1$}}
\newcommand{\vnu} {{\bf \nu}}

\begin{frontmatter}

\title{On the lower bound on the exchange-correlation energy in two dimensions}

\author[label1]{E.~R\"as\"anen}
\ead{erasanen@jyu.fi}
\author[label2,label3]{S. Pittalis}
\author[label2,label3]{C. R. Proetto\fnref{fn1}}
\author[label4]{K. Capelle}

\fntext[fn1]{Permanent address: 
Centro At\'omico Bariloche and Instituto Balseiro,
8400 S. C. de Bariloche, R\'{i}o Negro, Argentina}

\address[label1]{Nanoscience Center, Department of Physics, University of
  Jyv\"askyl\"a, FI-40014 Jyv\"askyl\"a, Finland}
\address[label2]{Institut f{\"u}r Theoretische Physik, Freie
  Universit{\"a}t Berlin, Arnimallee 14, D-14195 Berlin, Germany}
\address[label3]{European Theoretical Spectroscopy Facility (ETSF)}
\address[label4]{Departamento de F\'isica e Inform\'atica, Instituto
  de F\'isica de S\~ao Carlos, Universidade de S\~ao Paulo, Caixa
  Postal 369, S\~ao Carlos, S\~ao Paulo 13560-970, Brazil}

\begin{abstract}
We study the properties of the lower bound on the
exchange-correlation energy in
two dimensions. First we review the derivation of the bound and show 
how it can be written in a simple density-functional form.
This form allows an explicit determination of the prefactor of the
bound and testing its tightness. Next we focus on finite two-dimensional
systems and examine
how their distance from the bound depends on the system geometry.
The results for the high-density limit 
suggest that a finite system that comes as close as
possible to the ultimate 
bound on the exchange-correlation energy has
circular geometry and a weak confining 
potential with a negative curvature.
\end{abstract}

\end{frontmatter}

%[main text]
\section{Introduction} \label{Intro}
The lower bound on the quantum mechanical part of the Coulomb 
interaction energy, commonly known as the Lieb-Oxford (LO) 
bound~\cite{lieboxford81}, is a key concept in many-body physics. 
The bound is not only of fundamental importance for, e.g.,
analyzing the stability of matter~\cite{spruch}, but
it has also been used as a key constraint in the construction
of several exchange-correlation functionals within
density-functional theory (DFT), which is a standard
tool in electronic-structure calculations of atoms, molecules, 
solids, and various nanoscale systems~\cite{dft}. 
Recently, substantial efforts have been directed at testing the 
{\em tightness} of the original three-dimensional LO
bound~\cite{odashimacapelle07,odashimacapelle08} 
and in finding tighter forms~\cite{odashimatrickeycapelle08,prl}.

In view of the growing interest in two-dimensional (2D) systems,
the existence and properties of the 2D form of the
lower bound on the exchange-correlation energy are of immediate relevance.
In this paper we review the known form of the 2D bound and show
how it can be derived from scaling relations. 
We discuss the tight prefactor for the bound obtained from the 
properties of the homogeneous 2D electron gas (2DEG) in 
the low-density limit. We test this bound for finite 2D
systems, and, in particular, by varying the shape of the system 
for a simple two-electron system, we propose a set of general 
properties for a finite 2D system 
which is {\em as close as possible} to the
lower bound on the exchange-correlation energy.

\section{Two-dimensional bound}

Lieb, Solojev, and Yngvason~\cite{liebsolovejyngvason} (LSY) 
have rigorously derived a 2D form of the LO bound, which 
can be expressed in terms of the indirect part of the interaction
energy:
\begin{equation}
W_{xc}[\Psi]\equiv \left<\Psi|{\hat V}_{ee}|\Psi\right>-U[n]\geq - \, C\int d^2 r\,n^{3/2}(\vr),
\label{lo}
\end{equation}
where ${\hat V}_{ee}=\sum_{i>j}|\vr_i-\vr_j|^{-1}$ is the Coulombic
{\em e-e} interaction operator, $\Psi(\mathbf{r}_{1},...,\mathbf{r}_{N})$
is any normalized 2D many-body wave function, $n(\mathbf{r}) $ is the
corresponding density, and $U[n]$ is the classical Hartree energy.
For the prefactor LSY estimated $C \leq C_{\rm
  LSY}=192\sqrt{2\pi}\approx 481$.

Interestingly, the exponent $3/2$ in Eq.~(\ref{lo}) follows directly from
universal scaling properties of the {\em e-e} interaction~\cite{prl}.
Under homogeneous 2D coordinate scaling, $\vr\to \gamma\vr$ 
($0<\gamma<\infty$)~\cite{levyperdew93} 
the $(2N)$-dimensional
many-body wavefunction scales as $\Psi(\vr_1\ldots \vr_N)\rightarrow 
\Psi_\gamma(\vr_1\ldots \vr_N)=\gamma^{2N/2}
\Psi(\gamma\vr_1\ldots \gamma\vr_N)$,
preserving normalization. This produces
the number-conserving scaled density $n(\vr) \to n_\gamma(\vr) = \gamma^2
n(\gamma\vr)$. 
Further, $W_{xc}[\Psi] \rightarrow W_{xc}[\Psi_{\gamma}]
= \gamma W_{xc}[\Psi]$, since both the Coulomb
interaction and its Hartree approximation scale linearly.
Denoting the exponent in Eq.~(\ref{lo}) by $\chi$, we then
find the relation 
\be
\gamma W_{xc}[\Psi]\geq -\,C\,\gamma^{2(\chi-1)}\int d^2 r\,n^{\chi}(\vr),
\ee
so that consistency with Eq.~(\ref{lo}) gives $\chi=3/2$.

It was also conjectured in Ref.~\cite{prl} by the present authors that
the prefactor can be decreased to $C=1.96$ which corresponds to a
significant tightening of the original 2D bound with $C_{\rm
  LSY}\approx 481$.
The tightest bound, i.e.,
the lowest exchange-correlation energy corresponds to
the 2DEG in the low-density limit, and hence {\em all} other 2D
systems (including all real, finite systems) 
are energetically above this bound. However, 
it remains to be examined which type of a finite system is 
closest to the bound.

\section{Density-functional form of the bound}

Here we reformulate the bound in Eq.~(\ref{lo}) 
in terms of {\em density functionals}.
The right-hand side can be written 
in terms of the expression of the local-density
approximation (LDA) for the electronic exchange in 2D,
\be
E_x^{\rm LDA}[n] = -A\,\int d^2 r\,n^{3/2}(\vr),
\label{lda}
\ee
where $A=4\sqrt{2}/(3\sqrt{\pi})$. This formula
has the same scaling with respect to the density as Eq.~(\ref{lo}).
Note that Eq.~(\ref{lda}) is {\em exact} for the exchange energy of 
the 2DEG (constant $n$) by construction~\cite{rajagopal}. 
The left hand side of Eq.~(\ref{lo}) can be written
in terms of the exchange-correlation energy as defined
in DFT, i.e., $E_{xc}[n]\equiv W_{xc}[n]+T_c[n]\geq W_{xc}[n]$,
where $T_c$ is the difference between the many-body kinetic
energy and the (single-particle) Kohn-Sham kinetic energy
and it is always positive. Now, Eq.~(\ref{lo}) becomes
\be
E_{xc}[n]\geq \frac{C}{A} \; E_x^{\rm LDA}[n].
\label{const_lambda}
\ee
For any 2D system, we can now consider the density functional
\be
\lambda[n] = \frac{E_{xc}[n]}{E_x^{\rm LDA}[n]}\leq \frac{C}{A}
\label{LObound}
\ee
with $E_{xc}[n]=E_x[n]+E_c[n]$. As mentioned
above, the tightest 2D bound corresponds to the 2DEG
in the low-density limit which yields the {\em maximum}
value of Eq.~(\ref{LObound}), i.e., 
$\lambda_{\rm 2DEG}[r_s\rightarrow\infty]=1.84$,
where $r_s=1/\sqrt{\pi n}$ is the density parameter in
2D.

\section{Testing the bound}

\subsection{General remarks}

In practice, calculation of $\lambda[n]$ for finite systems and
thus testing the bound is seriously limited by the lack of
reference data for exact exchange-correlation energies
and exact densities. An exception is the
2D Hooke's atom, i.e., a parabolic (harmonic) quantum dot (QD)
with two electrons ($N=2$), for which analytic solutions are 
known~\cite{taut}, and which yields
$\lambda[n]\approx 1.55$ as the maximum value~\cite{prl}. 

It is also possible to approximate $\lambda$ from the
exact reference data solely for the total energy 
$E_{\rm tot}$. This requires exact-exchange (EXX) or
Hartree-Fock calculations to obtain the exchange energy,
which, for closed-shell systems, can be written as
\be\label{ex}
E_x[n]=-\int d^2 r \int d^2 r' \, \frac{\left|\sum_{i=1}^{N/2}\psi_i^*(\vr)\psi_i(\vr')\right|^2}{|\vr-\vr'|},
\ee
where $\psi_i$ are Kohn-Sham (or Hartree-Fock) orbitals. 
The correlation energy is then obtained as
$E_c=E_{\rm tot}-E^{\rm EXX}_{\rm tot}$. 
It should be noted that Eq.~(\ref{LObound})
becomes now an approximation due to using
both exact densities and EXX densities as the input
instead of consistently using only the exact densities.
Nevertheless, tests for few-electron parabolic and square-shaped 
QDs based on this strategy have led to values in the range
$1.1\lesssim \lambda[n] \lesssim 1.5$ (Ref.~\cite{prl}).
The results for QDs suggest that the 
largest value for $\lambda$ is obtained with $N=2$.

\subsection{Effects of geometry}

To consistently analyze the dependence of $\lambda[n]$
on the system geometry, we focus in the following 
solely on the limit $r_s\rightarrow 0$. This corresponds
to the noninteracting situation, since the kinetic
energy scales as $r_s^{-2}$ and the interaction energy as $r_s^{-1}$
(cf. the opposite limit where interactions dominate and lead to
Wigner crystallization~\cite{wigner}). 
Now, the correlation 
energy is zero, but Eq.~(\ref{LObound}) is still a well-defined 
quantity having a form
\be
\lambda[r_s\rightarrow 0] = \frac{E_{x}[n]}{E_x^{\rm LDA}[n]}.
\label{LObound2}
\ee
Furthermore, we set $N=2$ so that the exact exchange energy 
in Eq.~(\ref{ex}) can be calculated as a simple integral
over the density; in fact it is exactly minus half of the
Hartree energy. To find the density we solve the
Schr\"odinger equation for a noninteracting singlet state 
in the presence of an external confining potential $V(x,y)$. 
In the numerical procedure we use the {\tt octopus} code~\cite{octopus}.
Note that upon the condition that the potential scales
homogenously
with respect to the scaling parameter $\gamma$
(see above), we can choose 
any prefactor in the potential, i.e., any $\zeta$ in  $\zeta V(x,y)$,
in order to mimic the true interacting calculation in the limit 
$r_s\rightarrow 0$ corresponding to $n\rightarrow \infty$. 
This can be seen by considering $V(r)$ in the radial 
Schr\"odinger equation for $N=2$, where 
$\psi_i(r)\propto \sqrt{n(r)}$ ($i=1,2$). The form
of the equation under uniform coordinate
scaling $\vr\to \gamma\vr$ (see above) shows that
$V(r) \rightarrow \gamma^2 V(\gamma r) = \gamma^{2+\alpha}V(r)$,
assuming that $V(r)$ scales homogenously with $\gamma$. Hence, a value
for $\gamma$ can be found for any prefactor $\zeta$. In other words,
changing the prefactor is equivalent to scaling of the density.
And in fact, this is the case in all potentials given below, as all of
them scale homogeneously with respect to $\gamma$.
Moreover, Eq.~(\ref{LObound2}) is now independent of $\gamma$.

First we consider a circular QD defined by a confining
potential of the form
\begin{equation}
V_{\rm circular}(r)=|r|^\alpha,
\end{equation}
and a square-shaped QD defined by
\begin{equation}
V_{\rm square}(x,y)=|x|^\alpha+|y|^\alpha.
\end{equation}
Here the parameter $\alpha$ determines the ``steepness''
of the potential. 

Figure~\ref{deformation}
\begin{figure}
\begin{center}\leavevmode
\includegraphics[width=0.8\columnwidth]{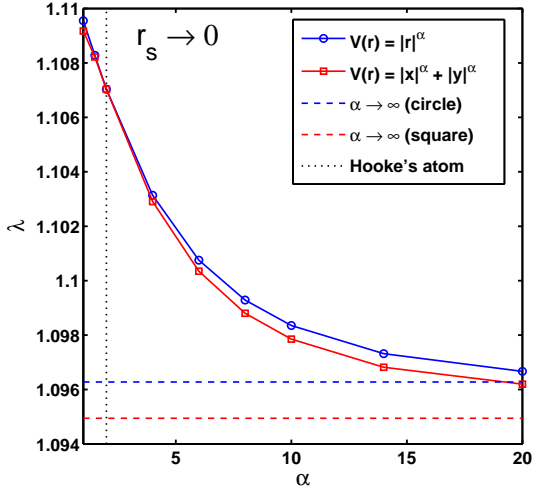}
\caption{(color online) Values for $\lambda$ in 
two-electron quantum dots with varying external
confining potential in the noninteracting limit.}
\label{deformation}
\end{center}
\end{figure}
shows $\lambda$ in the $r_s\rightarrow 0$ limit as
a function of $\alpha$. Obviously, the two potentials are 
the same when $\alpha=2$, when they actually correspond to the 
2D Hooke's atom (dotted line) with $\lambda\approx 1.107$.
At $\alpha>2$ we find monotonous decrease of $\lambda$, and
the limit $\alpha\rightarrow\infty$, which corresponds to
the hard-wall case, leads to $\lambda\approx 1.0963$ and $1.0949$
in circular and square QDs, respectively. Overall, the circular
potential gives higher values of $\lambda$ than the square one.

Interestingly, the highest values for $\lambda$ are obtained
at $\alpha<2$. In this regime, $\alpha=1$ corresponds to
a cone (pyramid) in a circular (square) QD. The numerical 
accuracy of our eigenvalue solver
limits the investigation to $\alpha=0.5$, which 
actually yields the largest $\lambda$ as seen in Fig.~\ref{deformation}. 
Therefore, we assume that the maximum value for $\lambda$ could
be found in a circular QD at $\alpha<0.5$, when the curvature
is negative, i.e., $d^2 V(r)/(dr^2)<0$.

In Fig.~\ref{ellipse}
\begin{figure}
\begin{center}\leavevmode
\includegraphics[width=0.8\columnwidth]{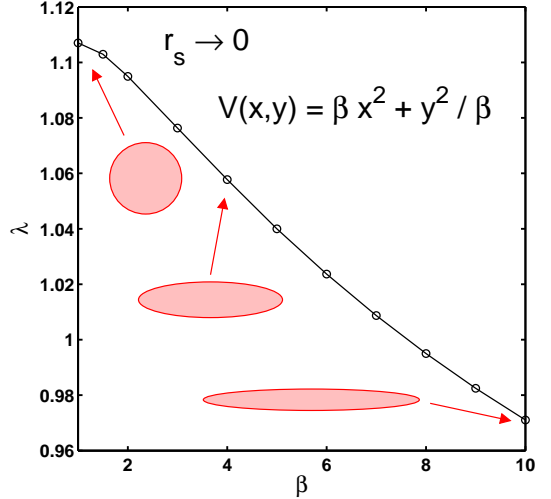}
\caption{(color online) Values for $\lambda$ in 
elliptic quantum dots in the noninteracting limit.
The shaded figures represent schematically the symmetry 
of the confining potentials.
}
\label{ellipse}
\end{center}
\end{figure}
we consider the $r_s\rightarrow 0$ limit in an elliptic 
confinement,
\begin{equation}
V_{\rm elliptic}(x,y)=\beta x^2 + y^2/\beta,
\end{equation}
where $\beta$ is related to the eccentricity, although
this is not a formal definition. Similar confinement has been
used in the QD studies in Refs.~\cite{ellipse1} and 
\cite{ellipse2}. We find that $\lambda$ decreases as 
a function of $\beta$, and the highest value can
be obtained in a parabolic system at $\beta=1$ (Hooke's atom).

\section{Conclusions and outlook}

Summarizing the results reported in the
previous section for finite systems,
we have found that 
in order to find a maximum value of $\lambda$ in the 
$r_s\rightarrow 0$ limit, the confinement potential should
be (i) circular instead of square or elliptic, and (ii) 
it should have a negative curvature.
Assuming that these geometric effects carry over to $r_s>0$, and
in particular to values for $r_s$ maximizing $\lambda(r_s)$,
and combining these findings with previous results for harmonic
systems with $N\geq 1$ and $r_s>0$ (Ref.~\cite{prl}), we advance the following
conjecture: The maximum value of $\lambda$ in {\em finite} 2D systems,
and hence the closest possible value to the ultimate bound on
the exchange-correlation energy can be obtained in a two-particle 
system having circular symmetry and a weak confining potential with
a negative curvature. 

Similar to what was previously observed in 3D~\cite{odashimacapelle07,odashimacapelle08,odashimatrickeycapelle08}, smaller
particle numbers in 2D produce larger values of the functional 
$\lambda[n]$ (Ref.~\cite{prl}). 
This should be compared with the behavior of the 
function $\lambda(N)$, which for any $N$ produces an upper limit on 
$\lambda[n]$ for all densities integrating to this $N$. In 3D, $\lambda(N)$ 
is known rigorously~\cite{lieboxford81,odashimacapelle07} 
to be monotonically increasing with $N$. We
expect this to be true also in 2D, but have not proved it. In any
case, the fact that the upper limit $\lambda(N)$ increases with $N$, while 
the actual value $\lambda[n]$ decreases is not a contradiction. It simply 
means that the LO bound becomes more and more generous as $N$ increases, 
and, conversely, tighter and tighter as $N$ decreases.

Overall, we conclude from the investigation of the present paper that, 
qualitatively and even semi-quantitatively, 2D systems behave similarly
to 3D systems with respect to the appropriate LO bound.

Furthermore, in view of the results for the 2DEG and 
2D Hooke's atom~\cite{prl} we may assume that a finite system 
with the largest $\lambda$ 
would have rather uniform density (except at the boundaries),
and $\lambda$ would be very close to the bound value $\lambda_{\rm
  2DEG}=1.84$. Although we cannot rigorously prove these
assumptions for the potential or for the density 
at large $r_s$, we hope that these results will
encourage similar geometric studies on finite systems 
where correlations are incorporated.
This would require using, e.g., the density-functional formalism
for strictly correlated electrons~\cite{paola}, or an inversion
scheme~\cite{reverse},
to reconstruct the exact external potential and the corresponding
many-body problem for a given density.

This work was supported by the
Academy of Finland, Deutsche Forschungsgemeinschaft, and
the EU's Sixth Framework Programme through the ETSF e-I3.
In addition, C. R. P. was supported by EC's Marie
Curie IIF (MIF1-CT-2006-040222) and K. C. by FAPESP and CNPq.

\end{document}